\documentclass[12pt]{article}
\usepackage{amsfonts,a4wide}

\makeatletter\relax
\openin\@inputcheck calrsfs.sty\relax%
\ifeof\@inputcheck
\else\closein\@inputcheck \relax%
\usepackage{calrsfs}
\fi\relax
\makeatother\relax

\font\msyms=cmsy10 scaled \magstep1
\def\pargsgn{{\msyms\char'170}}

\def\half{{\textstyle{1\over2}}}

\def\Ratio#1#2{{\textstyle{#1\over #2}}}

\def\mib#1{\hbox{\boldmath $#1$}}
\def\smi#1{\small$#1$}

\def\vc{{\vec{\mib v}}}
\def\en{{\cal E}}
\def\mom{{\vec{\cal P}}}
\def\fc{\vec{\hbox{\mib F}}}

\def\fir{\vec{\hbox{\mib F}}_{\rm in}}
\def\fext{\vec{\hbox{\mib F}}_{\rm ext}}
\def\Fm{{\vec{\cal F}}_{\rm mass}}
\def\Fi{{\vec{\cal F}}_{\rm mom}}
\def\Fgr{{\vec{\cal F}}_{\rm attr}}
\def\Fgyr{{\vec{\cal F}}_{\rm gyr}}
\def\Fcf{{\vec{\cal F}}_{\rm cf}}
\def\Fcor{{\vec{\cal F}}_{\rm Cor}}
\def\elf{{\vec{\mib E}}}
\def\mag{{\vec{\mib H}}}
\def\Com{{\vec{\cal X}}}
\def\Comm{{\vec{\cal Y}}}
\def\Chrg{\hbox{\mib e}}
\def\chrg{\hbox{\smi e}}
\def\Vpq{{\vec{\cal V}_{(pq)}}}
\def\Vpf{{\vec{\cal V}_{(p\phi)}}}
\def\Vqf{{\vec{\cal V}_{(q\phi)}}}
\def\mi{\mathrm{i}}

\def\Ct{C_{(\tau)}}
\def\Cs{C_{(\sigma)}}
\def\Cz{C_{(0)}}
\def\ct{c_{(\tau)}}
\def\cs{c_{(\sigma)}}
\def\cz{c_{(0)}}

\def\Op{\Omega_{(p)}}
\def\Oq{\Omega_{(q)}}
\def\Up{\Upsilon_{(p)}}
\def\Uq{\Upsilon_{(q)}}

\def\drp{{\partial\over\partial p}}
\def\drq{{\partial\over\partial q}}
\def\drf{{\partial\over\partial \phi}}
\def\drr{{\partial\over\partial r}}
\def\drt{{\partial\over\partial \theta}}

\def\P{{\cal P}}
\def\Q{{\cal Q}}

\def\bDer#1{{\mib D\over\mib\delta\mib\tau}#1}
\def\lDer#1{{\mib D#1/\mib\delta\mib\tau}}

\def\Dex{{\,{\cal D}\hskip 0.05ex}}

\newcommand{\poker}{%
\mathbin{\mbox{%
              \begin{picture}(5,5)
                            \put(0,1){\line(1,0){4}}
                            \put(4,1){\line(0,1){4}}
              \end{picture}}}}

\def\crosstimes#1#2{\big[#1\times#2\big]}

\def\ie{\hbox{\it i.\kern -0.12ex e.}}
\def\apriori{{\it a priori\/}}

\def\etc{{\it etc.}}
\def\cf{{\it cf.}}
\def\Cf{{\it Cf.}}
\def\eg{{\it e.g.\/}}
\def\rhs{{\sc r.h.s.}}

\def\lhs{{\sc l.h.s.}}
\def\ah{{\it ad hoc\/}}

\begin{document}

\begin{center}
\Large\bfseries 
Generalized Alignment  \\ of Gravitational Intencities\\
and Electromagnetic Strengths \\
in Kerr-Newman Space-Time
 \end{center}

\bigskip
\begin{center}
S.I. Tertychniy
\end{center}

\begin{center}
National Research Institute \\ of Physical-Technical
and Radiotechnical Measurements\\
(VNIIFTRI) \\
Mendeleevo, Moscow Region, 141570, Russia
\end{center}
\bigskip\bigskip\bigskip

\begin{abstract}
\noindent
It is shown that, in the case of Kerr-Newman space-time, the complex
electromagnetic strength $\elf+\mi\,\mag$ and analogous complex
intensity of the
gravitational field $\Fgr-\Ratio{\mi}{2}\Fgyr$
share the common complexified spatial direction.
 \end{abstract}
\bigskip

\section{Introduction}
\label{s1}

Considering
interrelation of the Newtonian physics and the special
relativity, it is instructive to mention that there arise no
obstacles standing in the way of the relativistic generalizing the
notion of a force in a fully coherent way, provided the basic theory
is re-formulated in a Lorentz-invariant form. In mechanics,
considering motion of a test particle, the force 4-vector
proportional to the particle 4-acceleration is used to obtain the
generalized relativistic definition of a force which exhibits a proper
non-relativistic limit. Accordingly, turning to the field theory, a
strength of a physical field, affecting the motion of particles which
carry the corresponding `charge', can be introduced on the base of the
relevant specific forces.
 %
(Electromagnetic field is the obvious and most important
instance where the above interpretation perfectly applies.)

A situation proves to be considerably more subtle when a
similar transition from the flat space-time to a curved one is undertaken. Of
course, considering motion of a classical test particle, both the
4-acceleration and the corresponding force 4-vector make sense
and might be utilized for the extending of the force dynamics
to the case of arbitrary curved
space-time. However the manifestations of a 4-force alone does not
exhaust the total effect the forces applied to a particle are
responsible for. Specifically, admitting the 4-force to
represent the resulting extrinsic influence to motion of a particle
(\ie\ the action of all the relevant physical fields living in the
space-time), the effect of the {\em gravitational field\/} proves to
be missed --- or, in the best case, characterized only indirectly and
highly incompletely through the specific influence of the other
fields affected by the gravitational distortion of the space-time
geometry.

Indeed, whenever the particle is free, \ie\ its worldline is geodetic
and the 4-acceleration identically vanishes, it often should not be
regarded to be unaffected by any
force at all (as the obvious example of the sucking in a black hole
demonstrates; less academic one is provided by the effect of
accretion studied in astrophysics, see, \eg, Refs.\ \cite{abs}).
Definitely, it
seems more physically adequate to consider a freely falling
particle to be, in a sense, `accelerated' (`dragged'?) in the
gravitational field described by the curved space-time. Accordingly,
an introduction of the corresponding gravitational force has to be
regarded a meaningful problem.

It is worth noting here that, generally speaking, there still exist just
opposite attitudes to the notion of a gravitational force.
Specifically, the latter may be regarded as an obsolete
speculative concept justified, in the best case, by the out-of-date
habit alone. Indeed, working exclusively
within framework of the 4-dimensional theory, a
gravitational force is obviously a superfluous notion.
However
one may also interpret gravitational force as an inherent
property of the corresponding physical field despite that the
particular state of the motion of observers heavily affects such a
characteristic and even may, in a sense, apparently eliminate it. The
example of electromagnetism, where the energy and the momentum
of the field
are the functionals over the observable electric and magnetic
strengths, may be regarded as an indirect argument in favor of the potential
plausibility of such a relationship.

Not needing to predetermine a final judgement on the above dilemma,
it is anyway worth analyzing the possibility of the consistent
extending the concept of gravitational force to the case of a
strong space-time curvature. It is especially important to follow a
way strictly concordant with the weak field limit and to operate within
physical framework (as opposed to formal
mathematical considerations).

It is worth noting that investigations in the fields
bearing a relation
to the
issue mentioned are of a comparatively long history. The latter is
briefly reviewed in Ref.\ \cite{grqc} where the extended list
of references is given, see also Refs.\ \cite{BCJ}, \cite{rev} and
references therein.

Nevertheless
the problem seems to be not properly cleared up.
The goal pursued in the present work is to propound
a consistent
method of the description of the `true' gravitational field
which suits for arbitrarily strong
gravitational field and simultaneously reveals the proper Newtonian limit.
We shall see that it entails some new insight into the
problem of description of
observable manifestations of the gravitational field.

 Generally speaking, the realization of the dynamical description of
the gravitational field (\ie{}, here, the one referring to the
concepts of forces and field intensities) is more intricate than the
analogous treatments of the other classical fields, the
electromagnetic one being a primary working object for a collating,
of course. Concerning the physics, the two reasons should be
mentioned in this respect.

The first intrinsic cause of troubles is the well known
unavoidable co-existence of gravitational forces in conjunction with
inertial ones. Accordingly, attempting to describe the dynamical
properties of the gravitational field itself, one has to be able to
unambiguously distinguish the manifestations of the gravity from ones
of the inertia. At the same time, within framework of the general
relativity, the meaning of the separating the gravitational and
inertial phenomena remains to be not satisfactorily understood. A
conventional principle of the distinction of gravitational and
inertial forces has not been clearly formulated; the formal
analogues, mostly of mathematical origin (or even looking sometimes
like terminological tricks), being used in practice instead.
Sometimes, realizing the explicit splitting of forces into
gravitational and inertial constituents, such a procedure is regarded
as an artificial trick carried out after \ah\ fashion for the sake
of convenience or a qualitative visualization of the covariant
equations at most.

There is however a standpoint supposing an
intrinsic character of the separation of the
gravitational field and inertia manifestations
within their combined `gravitational-inertial bundle'.
Although
the broadly interpreted principle of equivalence,
see Ref.~\cite{MTW},
states that
the gravity and the inertia reveal themselves in local effects in
indistinguishable ways (\cf\ Ref.~\cite{RP}), it may nevertheless be
supposed that there exist some their aspects which make them
physically non-equivalent.

Specifically, there exist today no reasons (at least, for a theorist)
to doubt that gravitational field possesses some own energy
`separated', in a sense, from the own energy of gravitating bodies.
Moreover, this energy is able to be born across the space in a form
of gravitational waves and, then, to to be partially handed over to
remote physical objects (\eg, to a gravitational antenna). On the
other hand, the inertia manifestations, heavily dependent on the
motion of observers, surely cannot be associated with any such a long
scale {\it energy transfer\/} detectable by an independent observer.
It may be therefore argued that the own energy associated with the
`gravitational-inertial bundle' given is to be differently connected
with its two `constituents', no matter this relation is not clear
yet.

It is worth noting that the only outcome of the above (perhaps not
perfectly flawless) speculation which we
would like to refer to
is a certain evidence of a potential
physical meaningfulness of attempts to separate the
manifestations of the `active' gravity from `superficial'
effects due to the `passive' inertia.
There are no evident reasons why one should \apriori\ regard such a
procedure as a mathematical exercise at most.

The next source of difficulties,  standing in the way of description of
the `pure' gravity in terms of the measurable quantities, is the
necessity to refer any observable force to some
`observational platform' --- a frame of reference. Unfortunately, in
the case of a curved space-time, the concept of a frame of
reference, formulated in a way concordant with the flat space-time
limit, cannot still now be regarded, on the whole, sufficiently
definite and clear.
In particular,
the status of a successor of inertial frames of
reference, constituting the foundation of the special relativity,
remains to be not properly understood yet%
.

There exists however an opportunity to evade the surmounting
the above problems in their generic posing. To that end, the present
work focuses on a particular case {\it intermediate\/}
between the flat and generic curved space-times where a sufficient
symmetry (whose extreme degree characterizes the flat
space-time) and arbitrarily strong curvature (associated with
non-ignorable gravitational field) are combined. The
subject of our analysis --- Kerr-Newman
space-time~\cite{KN} --- is one of the most important
solvable models in the gravity theory.
This case encompasses also
Kerr, Reissner-Nordstr\"om and
Schwarzschild fields (see Ref.\ \cite{Cat}) as particular limits, the
results of the work holding true {\it mutatis mutandis\/} for them as
well.

The symmetry of the model picked on immediately gains us the well
known natural way of the introduction of the reference platform
(`non-inertial frame of reference') possessing a clear
interpretation in physical terms.
%
Accordingly,
the theoretical tools used in the present work for the referring the
spatial and temporal relations in Kerr-Newman space-time in terms of
observable quantities (the so called apparent space and its
derivatives, see section \ref{s2}) are based just on a remnant the
model inherits from the Minkowski space-time.

As opposed to the usual tendency, we do not strives to make the
model maximally general. Instead the goal is pursued to limit
ourselves with the only tools whose physical meaning is absolutely
transparent, and to minimize the making use of any
additional independent assumptions. Attaining it, there
appears an opportunity to more or less directly apply the standard
means used in similar situations within framework of the flat
space-time theory.
As a matter of fact, this element of the method possesses an
unambiguous physical interpretation and proves to be a strictly
unique one (in frame of the problem posing assumed).

Further, the standard way of the introducing of observable forces
acting to test particles is followed. It allows one to define the
corresponding static strengths (intensities) of physical fields
living in Kerr-Newman space-time, yielding ultimately the
completely plausible results. In particular, the observable
intensities characterizing the gravitational-inertial manifestations
and the observable strengths of the electromagnetic field immediately
result.

Furthermore, by virtue of the model transparency, a straightforward
speculation enables us to feel about for a method of the separating the
gravity and inertia forces and to determine the true intensities of
the gravitational field --- attractive (gravitoelectric) and
gyroscopic (gravitomagnetic) ones.
 The close connection of the gravitational field to the space-time
curvature is the guiding relationship utilized.

It is shown in particular that in
the case of Kerr-Newman field the standard complex combination
of the electric and magnetic strength vectors and analogous complex
combination of the true gravitational intensities are
{\it proportional\/} (over $\mathbb{C}$) that may be
referred to as the {\it generalized alignment of the gravitational
and electromagnetic fields}. To the best our knowledge,
such a relationship is revealed for the first time.
Another surprising implication is a somewhat unexpectable form of the
relationship of the intensities of the gravitational field and the
curvature: we show that the former may not be regarded as a plain
form (such as, \eg, a projection) of the latter. (For more details
see section \ref{sum}.)


As it was mentioned above,
the related problems, concerning in particular the concepts of
gravitational and inertial forces, have been treated, one way or
another, in a number of works from the positions some of whose are of
definite similarity to our one while others substantially differ.
For the sake of easier comparison, we list in short the main
approaches presented in the literature.

First of all
it should be noted that the elements of the method used in the
present work can be found in Ref.\ \cite{LL} (see in particular
therein the discussion of {\sl Problem~1} following
\pargsgn{}88). In the both cases the approaches are based on the
analysis of the elementary measurement procedures and follow rather a
physical intuition than mathematical likenesshood.

A semi-heuristic approach applied in Refs.~\cite{FF1}-\cite{FF3}
for the introduction of the true gravitational force is based on a
plausible assumption concerning its `Lorentz-like' transformation
rule. The corresponding results, describing the attractive
gravitational and centrifugal forces in Schwarzschild space-time and
Reissner-Nordstr\"om space-time (see in particular Eqs.\ (12a), (12b)
in \cite{FF2}), agree with the particular cases of our formulae (see
section \ref{s8}). However the gyroscopic forces (gravitomagnetic and
Coriolis) are not taken into account there.

The extended series of works \cite{ab1}-\cite{ab21} follows substantially
different ideas, basing on the notion of the so called `optical
geometry' \cite{ab3}. The physical background lying beyond the
latter notion is exhibited in Ref.\ \cite{ab9}. It relies first of all
on a special regarding of some globally synchronized rescaled
`universal time' which is obtained from the proper time element by
means of some conformal transformation, see \cite[page 7]{ab9}. This
implies the regarding of the role of physical time distinct from one
assumed in the present work, see section \ref{s2}, statement (B).
The possibility of global synchronization ensured by the conformal
transformation is not a primary mandatory property of a time since the
measuring of the latter
is a local procedure.
From our point of view the global `universal time' introduced in
\cite{ab9} is not actually a {\em time\/} since it is not measured by
any clock (showing proper time intervals along the clock' worldline).
The non-constant rescaling of
readings of watches,
used for explanation of the meaning of conformal transformation,
means much more than a mere choice of the
`local' units of temporal duration.
Further, the optic geometry treats trajectories of photons (light
rays) as genuine straight lines. However it seems to be not a quite
plausible conjecture. Specifically, by virtue of the equivalence
principle, it is unreasonable to suppose that the light is not
affected by the gravitational field. Indeed, the light is a `high
frequency limit' of the material object possessing the energy and,
hence, the gravitational mass --- periodic electromagnetic waves. The
properties of light indeed allows one to obtain the absolute measure
of a velocity {\em magnitude\/} (gravitational field is not able to force
light signal to move faster or slower because its speed is maximal,
anyway) but does not yield the standard of an invariable {\em
direction}, the gravitational deflection of light and gravitational
lensing being the well known relevant observable effects connected
with variation of light ray direction. It is also worth noting that
the approach in question involves a number of independent statements
(claiming, for example, that (citing) ``in static space-time
gravitational force is velocity independent'', \cite[page 943]{ab13},
\etc) which seem to be not convincingly motivated from the physical
point of view. Resuming, it is stated in Ref.\ \cite[page 734]{ab8} that
(citing) ``the rotational effects $<$\dots$>$ can be properly
understood only after a fundamental revision of the very concept of
centrifugal force''. The present work demonstrates however that the
traditional regarding of the latter quite suffices and this is a nice
sign since we should abstain from superfluous revisions of basic
concepts until this is being allowed by the logic of problem.

Next, in Ref.\ \cite{sem2} the method of interpretation of particle motion
in Kerr space-time in terms of forces is suggested. The observers
picked on are so called zero-angular-momentum ones (ZAMOs, see also
Ref.\ \cite{sem1}) which are (citing) ``standardly referred to as
proper generalization of the Newtonian non-rotated rest ones", some
specific hypothesis on the properties of centrifugal force in such a
frame being asserted. However the resulting description of the
particle dynamics proves to entail improper non-relativistic limit.
Indeed, Eqs.\ (64)-(67) of Ref.\ \cite{sem2} (with $M=0$) state that
centrifugal inertial force may exist while Coriolis one identically
vanishes.
In the further works by the same author \cite{sem3}-\cite{sem5} the
general covariant equations of the force balance in arbitrary
space-time were suggested (including Kerr space-time as a particular
example). The physical motivation of the method seems to be not quite
sufficient however. For example, the gravitational force is chosen to
coincide, up to some scalar factor, with the 4-acceleration of an
observer. The only restriction imposed on the frame is the claim for
observers' worldlines to be hypersurface-orthogonal
(`hypersurface-orthogonal observers', HOOs, see \cite{sem3}). Thus
the model suggested admits the existence of gravitational forces even
in the case of the flat space-time, provided the observers'
worldlines are not geodetic. From a physical point of view such a
development of the conventional concept seems to be not truly
plausible.

The general approach widely used for the re-casting the equations of
motion of test particles to the form of a force balance condition
is based on the geometric projections of 4-dimensional
characteristics to the directions parallel and orthogonal to
observers' 4-velocities. Alternative but similar method utilizes
space-like space-time slicings within framework of the 3+1 splitting
method.
 These  are
discussed in details in Ref.\ \cite{rev} and summarized recently in
Ref.\ \cite{grqc}. In Ref.\ \cite{BCJ} the relative gravitational force
is defined via the relative 4-acceleration of the two families of
observers. They play the role of the frames of reference and are
allowed to be chosen without substantial restrictions. The
gravitational force acting to test particles is defined in terms of the
spatial projection of the Fermi-Walker derivative of the particle
4-momentum while the notion of inertial forces is not drawn in at all. In
Ref.\ \cite{BRS} Kerr-Newman metric is considered. The approach in
question does not attempt to establish a relation of the
mathematical formalism developed to the corresponding procedures of
elementary measurement acts and observations which would confirm
the interpretation of the formal relationships derived in physical
terms.
Similar method based on the space-time slicing is utilized also in Ref.\
\cite{Para}. The gravitational force is introduced in framework of
the 3+1 splitting method. The projected equations of the test
particle motion are re-interpreted basing on the analogies with the
flat space-time dynamics. No inertial forces are taken into account.

In Ref.~\cite{QS} the `pseudo-Newtonian' gravitational force is
defined by means of a sort of `integrating' of the tidal gravitational
force. The latter is expressed via the projection of the Riemann tensor.
However the role of inertial forces is not allowed for.

Comparing with majority of the approaches mentioned, our one is
distinctive by (i) its close connection to the physical origin of the
problem, focusing on observable quantities and models of elementary
measurement procedures, and (ii) the utilizing as far as possible the
approaches and concepts developed within framework of the flat
space-time theory with minimal drawing in additional postulates. Our
results do not pretend at the current stage to any degree of
generality (in particular, by virtue of a substantial role of the
symmetries stipulated). Nevertheless a cogency of their
interpretation is the important outcome which could serve a ground
for the further consistent extending of the concept of gravitational
force to more general physical situations. A general theory
has to reproduce them in the corresponding particular cases.

\section{Spatial relationships in Kerr-Newman field from standpoint of
uniformly rotated observers}
\label{s2}

Let us consider the line element of Kerr-Newman space-time (or Kerr
space-time) in Carter' representation (see
Refs.~\cite{Car},\cite{Pl}):
\begin{equation}
-\mib{d}s^2
=(p^2+q^2)
\left({\mathrm d
         p^2\over\cal P}+{\mathrm d
q^2\over\cal Q}\right)+
{1\over p^2+q^2}
\left(
\P(\mathrm d\tau+q^2 \mathrm d\sigma)^2
-\Q(\mathrm d\tau-p^2 \mathrm d\sigma)^2
\right).
                                      \label{eq1}
\end{equation}
Here $p,q,\tau,\sigma$ are the coordinates,
$\P=\P(p), \Q=\Q(q)$ are smooth functions.
For simplicity, we
require $\P, \Q$ to be
positive in the connected space-time region which will be considered.
Since Kerr metric
is recovered when
\begin{equation}
\P=\P_{\rm K}(p)=a^2-p^2,\quad
\Q=\Q_{\rm K}(q)=q^2-2mq+a^2,
                                        \label{eq2a}
\end{equation}
while for the Kerr-Newman solution
\begin{equation}
\P=\P_{\rm KN}(p)=\P_{\rm K}(p),\quad
\Q=\Q_{\rm KN}(q)=\Q_{\rm K}(q)+\Chrg^2,
                                         \label{eq2b}
\end{equation}
(see Ref.\ \cite{Pl}),
$\P$ and $\Q$ are actually positive, provided
$m>a>0$, $q>m+\sqrt{m^2-a^2}$, $-a<p<a$.
The symbols $m$, $a$ and $\Chrg$ denote real constant parameters:
the mass, the specific angular momentum, and the electric charge of the
`central' source, respectively.
(For the choice of the physical units assumed
the light velocity {\mib c} and the gravitational constant are equal to
the unity.)
The space-time region discriminated by the above conditions is
adjacent to the asymptotically flat infinity. (The diverging observed
whenever $p\rightarrow\pm a$ can be shown to represent an
apparent peculiarity reflecting the local fault of the coordinate
system alone).

We shall need also the potential of the
electromagnetic field filling Kerr-Newman space-time. Its covector
equals
 \begin{equation}
{\cal A}={\Chrg\cdot q
          \over p^2+q^2}\:
                          (\mathrm d\tau-p^2 \mathrm d\sigma).
                                                \label{empot}
\end{equation}

It is worth noting that Boyer-Lindquist' ({\sc bl}) representation
\cite{BL} of Kerr-Newman metric is derived from
Eqs.~(\ref{eq1})-(\ref{empot}) by means of the local coordinate
transformation
 \begin{equation}
p=a\cos\theta,\quad
q=r,\quad
\sigma=-a^{-1}\varphi,\quad
\tau=t-a\varphi.
                                                \label{BLc}
\end{equation}
It proves to be more convenient for us to carry out all the
intermediate work making use of Carter' coordinates since they enable
one to gain advantage of a remarkable symmetry of the `radial' and
`angular azimuthal' coordinates, $q$ and $p$, respectively, which
Kerr-Newman metric possesses and which is manifest just when using
Carter' representation. On the other hand the interpretation of the
physical relationships implied is usually more manifest in terms of
{\sc bl} coordinates. In particular Carter' coordinates obviously
fail in the case of the vanishing of the angular momentum parameter
$a\rightarrow 0$.

\medskip

We shall consider the family of (formally structureless
point-like) observers
whose worldlines are described by equations
 \begin{equation}
  \quad p,q\mbox{\ are constant,\ }
  \sigma=-a^{-1}(\phi+\omega t),\quad\tau=(1-a\omega)t-a\phi,
  \mbox{\ $\phi$ being a constant.}
                           \label{eq3}
\end{equation}
In Boyer-Lindquist coordinates these read
 $
\varphi=\omega t+\hbox{const},\enskip
r,\theta\hbox{\ are constant}
$
(\cf\ Ref\ \cite{RP}).
In the flat space-time limit the observers reduce to uniformly rotated ones.
In Eq.\ (\ref{eq3}) the {\sc bl} coordinate $t$ may
be regarded as a free parameter on the curves of the congruence. The
parameter $\phi$ (identified modulo $2\pi$) together with (constant)
$p$ and $q$ label every individual observer. $\omega$ is the
predefined constant (the {\it formal frequency}) associated with the
whole congruence and characterizing the rate of the steady rotation
of the family of observers as a whole.
(One should not attach {\it a priori\/} a direct physical meaning to
the numeric values of
$\omega$, noting however that
in the case $\omega=0$ the observers may be
regarded non-rotating with respect to the static flat asymptotic.)

Let us mention that,
referring to a congruence of worldlines of point-like observers, we
deal with the realization of the `observational platform' which was
named in Ref.\ \cite{grqc} a ``threading point of view''.

It is worth noting that the norm of the vector field
\begin{equation}
U 
=
(1-a\omega){\partial\over\partial\tau}
-
{\omega\over a}{\partial\over\partial\sigma}
                                                        \label{Udef}
\end{equation}
tangent to the congruence (\ref{eq3})
amounts to
\begin{eqnarray}
  \vert\vert
  U
  \vert\vert^2&=&{\P\Q\over a^2(p^2+q^2)} [\mib{P}-\mib{Q}],
                                                       \nonumber
                                                                \\
\mbox{\llap{where\ }}
&&
\mib{P}=\mib{P}(p)=
{\displaystyle(a-\omega(a^2-p^2))^2\over\displaystyle\P
^{\vphantom I}
},\enskip
\mib{Q}=\mib{Q}(q)=
{\displaystyle(a-\omega(a^2+q^2))^2\over\displaystyle\Q
^{\vphantom I}                                         }. \nonumber
\end{eqnarray}
Thus
the congruence~(\ref{eq3}) is timelike, provided
 \begin{equation}
\mib{P}-\mib{Q}>0.
                           \label{eq4}
\end{equation}
Giving suitable $\omega$, we restrict consideration to the space-time
region (assuming it to be non-empty) where the above timelikeness
condition is fulfilled.

Having made these preliminary remarks, let us consider a pair of
close observers, associated, say, with the
(fixed) parameter values $\phi,p,q$ and
 $\phi+\mathrm d\phi,p+\mathrm d p,q+\mathrm d q$,
respectively.
Analyzing the exchange by light signals,
it is straightforward to show
that the distance $\mib{d}l$ separating these two
infinitesimally close observers amounts to the square root of
the quadratic form
 \begin{eqnarray}
\mib{d}l^2=
(p^2+q^2)
\left\{
{\mathrm d p^2\over\cal P}+{\mathrm d q^2\over\cal Q}
+{\mathrm d\phi^2\over\mib{P}-\mib{Q}}
\right\}.
                           \label{eq5}
\end{eqnarray}
It is positively defined by virtue of the condition~(\ref{eq4}).

From a mathematical standpoint the symmetric tensor
determining the quadratic form (\ref{eq5}) can be obtained
as a pull-back of the space-time tensor
$
h_{\mu\nu}=
$ $
(-g_{\mu\nu}
$ $
+ (U_\lambda U^\lambda)^{-1}
$ $
U_\mu U_\nu
)
$
with respect to the map
 $(p,q,\tau,\sigma) \rightarrow
(p,q,\phi=-\omega\tau -a(1-a\omega)\sigma)$.
Here $g_{\mu\nu}$ denotes the components of the metric tensor
determining the line element $\mib{d}s^2$, Eq.\ (\ref{eq1}),
$U_\mu$ denotes the covariant components of the vector field
(\ref{Udef}).
(\Cf\ the derivation of Eqs.\ (84,6), (84,7) in Ref.\ \cite{LL}.)
Thus the metric (\ref{eq5}) may be also regarded as minus the
projection of the space-time metric onto the infinitesimal 3-spaces
orthogonal to the observers' worldlines.

It has to be noted that Eq.\ (\ref{eq5}) is in fact a
quantitative realization of the following statements
which are not usually explicitly formulated:
 \begin{itemize}
\item[(A)] The speed of light detected by an arbitrary observer in
his/here {\it sufficiently small neighborhood\/} coincides with the
universal constant $\mib c$ independently of the features of
observer motion and the state of the ambient gravitation field.
\item[(B)] The time
measured by any observer coincides with the interval (the proper
time) $\mib{c}^{-1}\int\sqrt{\mib{d} s^2}$ accumulated along
his/here worldline.
 \end{itemize}
It is also implied that observers determine their mutual distance by
means of a local light ranging. This method is interpreted within
framework of the geometric optic, light rays being represented by
(infinitesimal) segments of null geodesics.

It is important that, referring to an arbitrary small neigbourhood of
an observer, the statement A stipulates the possibility to make a path of
the light signal arbitrarily short that ensures the necessary
relaxing of the role of the deflection of the probing ray due to all
the causes (except of a corner in the point of reflection from a
`target'). Essentially, the claim (A) expresses the exclusive role of
light signals for the ensuring the relativistic consistent measure of
a length.
This is a commonplace under the conditions corresponding to the special
relativity. Moreover, in practice, the standard of length is
established as a fixed number of wavelengths of a definite
transition line
in cesium.
All the more, in the case of essential role of
the acceleration or the gravitational field, there may exist no
relativistic means to introduce the notion of a length other than to
assume the invariance of the local velocity of light. In principle,
such an assumption might lead to physical inconsistencies (whose
actual existence is to be verified with the help of observations
analogous to Michelson' experiment) but in such a case this is the
notion of a local infinitesimal length which would fail to be
physically meaningful.

The important property of the length element (\ref{eq5}) is its {\it
independence\/} on a particular moment of measurement. The latter
fact is a trivial consequence of the obvious property of the
congruence (\ref{BLc}): it consists exactly of the orbits of a
space-time isometry group. Indeed, the vector field (\ref{Udef}) is
evidently the Killing field for the metric (\ref{eq1}) whose
coefficients depend on neither $\tau$ nor $\sigma$. Hence the
`development towards the future' of the observers is geometrically
the space-time isometry motion. Thus, from their standpoint any
invariant characteristic of the spatial geometry observed is
necessarily invariable with time. The infinitesimal distance is just
an instance of such a characteristic.

Thus it can be stated that
any two nearby observers from the family chosen may be regarded
{\it mutually immobile}. On an equal footing it may be also said that
they are in fact `immersed' in the {\it static Riemannian space\/}
whose metric is represented in the coordinate chart $\{p,q,\phi\}$ by
the expression (\ref{eq5}).


Hereinafter, for the sake of convenience of references, this
3-dimensional space will be named the {\it apparent
space\/}. By virtue of the universality and the very meaning of the
above claims~(A) and~(B), the apparent space encodes {\it all\/} the
local spatial relationships leaned on the notion of a {\it distance\/}
which are available for the observers under consideration.

It is worth emphasizing that the latter claim is based on a
theoretical model of the physical observation carried out by
observers and involves no additional conjections.

It should exist no serious objections against the statement that
the apparent space is a
really existing physical-geometrical entity and that particular local
observations may (and have to) refer to it for the {\em concordant
joining\/} local results into a unified picture.

This way, `solidifying' the observational platform, we come at a
qualitative situation which is characteristic of the special
relativity and which can be consistently treated by the
straightforward methods analogous to ones applied in the case of the
flat space-time.

\section{Kinematic of a test particle in apparent space}

Let us consider a test particle moving in accordance with the
equations $x^\alpha=x^\alpha(\nu)$, where $\nu$ denotes the proper
time along the particle worldline. Given the notion of
3-dimensional apparent space, the vector of
 velocity $\vc$, the vector of momentum $\mom$
(whose symbol should not be mixed up with the function $\P= \P(p)$)
and the energy $\en$ of the particle can be determined.
The
above characteristics of particle motion are expressed in terms of
invariant geometrical objects (scalars and vectors) {\it over
the apparent space}.

Specifically, in physical terms, the vector of speed is to be defined
as an infinitesimal displacement of the particle divided to the (also
infinitesimal) duration of the corresponding motion which
is measured by the observer being in the point occupied initially by
the particle. It is worth mentioning that the time interval of the
particle motion is not \apriori\ referred to any prescribed `timelike
coordinate' or the 3+1 space-time splitting. Rather it is defined by
means of the modeling of the exchange of light signals (\ie{} the
local light ranging) and is based on the above claims~(A) and (B)
which just suffice for this purpose, provided the observers carrying
out the measurements are properly specified.

 Further, knowing the speed, the energy and the momentum of a
particle follows from the standard local relativistic equations
generalized to the case of non-flat spatial ground (the apparent
space). This way, it is straightforward to show that
 \begin{eqnarray}
\en&=&
\sqrt{p^2+q^2\over\P\Q}\cdot
{
\omega\Cs-a(1-a\omega)\Ct
\over\sqrt{\mib{P}-\mib{Q}}
},
                                        \label{endef}
  \\
\mom&=&
{1\over p^2+q^2}\left\{
\epsilon_{(p)}
R_{(p)} 
\drp
+
\epsilon_{(q)}
R_{(q)} 
\drq
                \right.
                       \nonumber\\
&&
\hphantom{{1\over p^2+q^2}\left\{\vphantom{1\over p^2+q^2}\right.}
-\left[ (\Ct p^2+\Cs)
        {a-\omega(a^2-p^2)\over\P}
          \right.
                 \nonumber\\
&&
\hphantom{1\over p^2+q^2}
\left.\hskip 3.9 ex
 \left.
        +
        (\Ct q^2-\Cs){a-\omega(a^2+q^2)\over\Q}
 \right]\drf\right\},
                                                \label{eq6}
                                               \\
\mbox{\llap{where\ }}
 R_{(p)}&=&
       \sqrt{\P(\Cz-\mu^2p^2)-(\Ct p^2+\Cs)^2},
                                              \nonumber\\
 R_{(q)}&=&
       \sqrt{(\Ct q^2-\Cs)^2-\Q(\Cz+\mu^2q^2)}.
                                              \label{Rpq}
\end{eqnarray}
Here $\mu$ is the (constant) mass of the particle at rest. The
symbols $\Ct,\Cs,\Cz$ may be interpreted as the notations of
the following expressions:
\begin{eqnarray}
\Ct&=& {\mu\over p^2+q^2}
\left[\P\left({d\tau\over d\nu}+q^2{d\sigma\over d\nu}\right)
-\Q\left({d\tau\over d\nu}-p^2{d\sigma\over d\nu}\right)
\right],
                                        \nonumber
        \\
\Cs&=& {\mu\over p^2+q^2}
\left[q^2\P\left({\mathrm d\tau\over \mathrm d\nu}
+q^2{\mathrm d\sigma\over \mathrm d\nu}\right)
+p^2\Q\left({\mathrm d\tau\over \mathrm d\nu}
-p^2{\mathrm d\sigma\over \mathrm d\nu}\right)
\right],
                                        \label{eq7}
        \\
\Cz&=&\half\mu^2
\left[
(p^2-q^2)
+(p^2+q^2)^2
\left(
      {1\over\P}\left({\mathrm d p\over \mathrm d\nu}\right)^2
-     {1\over\Q}\left({\mathrm d q\over \mathrm d\nu}\right)^2
\right)\right.
                                        \nonumber
              \\
&&
\hphantom{\half\mu^2\left[ \vphantom{
  \left[q^2\P\left({\mathrm d\tau\over \mathrm d\nu}
+q^2{\mathrm d\sigma\over \mathrm d\nu}\right)\right]
                               }\right.}
\left.
+\P
\left({\mathrm d\tau\over \mathrm d\nu}
+q^2{\mathrm d\sigma\over \mathrm d\nu}\right)^2
+
\Q\left({\mathrm d\tau\over \mathrm d\nu}
-p^2{\mathrm d\sigma\over \mathrm d\nu}\right)^2
\right].
                                        \nonumber
\end{eqnarray}
At the same time
these are none other than the {\it geodesic integrals\/} for the
metric~(\ref{eq1}), \cf\ \cite{Pl}. Thus, claiming
$\Ct,\Cs,\Cz$ to be constant (together with the condition specifying
$\nu$ as the proper time), one obtains the complete set of the first
integrals of the geodesic equations for the metric~(\ref{eq1}). The
symbols
$\epsilon_{(p)}=\mathrm{sign}(\mathrm d p/\mathrm d \nu),
\epsilon_{(q)}=\mathrm{sign}(\mathrm d q/\mathrm d \nu)$,
involved in Eq.~(\ref{eq6}),
equal to $+1$ or $-1$. They however are not
the true constants because the sign of each of them is reversed in the
points where the argument of the corresponding square root,
either $R_{(p)}$ or $R_{(q)}$,
vanishes (and where the value of $\epsilon_{(*)}$ does
not matter), see Eqs.\ (\ref{Rpq}).

It is useful to notice in conclusion that the
momentum vector admits the alternative representation
   $$
\mom=
\mu\left(
   {\mathrm d p\over\mathrm d  \nu}\drp
+{\mathrm d q\over\mathrm d  \nu}\drq
+{\mathrm d \phi\over\mathrm d  \nu}\drf
\right),
   $$
\ie, it is a pull forward of the vector $\mu\cdot\partial/\partial\nu$
tangent to the particle trajectory.
This is not surprising since by virtue of definitions the direction
of momentum coincides with one of infinitesimal displacement of the
particle in the apparent space.

\section{The observable dynamics of test particles}
\label{s4}

Given the particle momentum, the problem of the determination of its
{\it physical variation rate\/} can be posed. For
definiteness we denote the latter as $\lDer{\mom}$
($\mib{\delta}\mib{\tau}$
means the infinitesimal {\it physical time interval\/} and should not
be mixed up with the variation of the coordinate $\tau$). The very
realization of the operator $\lDer{}$ has to be based on physical
arguments which are discussed below.

Let us consider
the generalized relativistic
version of the Newton law
$$
\fc=\bDer{\mom}.
$$
Essentially, this equation simply allows one to introduce the
notion of {\it observable force\/} acting to a test
particle. The latter may be regarded as the `cause' of {\it
observable acceleration\/} of the particle {\em in the apparent space}.
This line may be further continued pursuing the goal of the rigorous
introduction of {\it observable strengths\/} (intencities) of
physical fields including gravitational one. Accordingly, one may
interpret the forces as the result of the action of the fields to the
corresponding `charges' the particle possesses.

The basic relativistic recipe of the determination of the rate of
vector variation is of course well known and may sound as follows:
Let the two nearby points on the particle trajectory be chosen, the
first of them coinciding with position of the observer carrying out
the measurement. Further let the vector `dragged' with the particle
(speed, momentum, \etc) be registered in the two points mentioned,
and the corresponding difference be found. Dividing it to the duration of
the particle motion, the desirable `rate of vector variation'
follows.

Concerning the
time of the particle motion, it is determined by the
observer, occupying the point of the initial particle position. The
basic setup of its measurement is standard: the observer emits a
series of light signals towards the second observer which is situated
in the point of the future final particle position and registers the
light signal reflected by the second observer at the moment when the
particle arrives at the his/her point. Then, from the standpoint of the
first observer, the middle between the events of the emission and
the reception for that light signal is just the moment of the
particle arrival at the final point (precisely by virtue of the statement
(A) formulated in section~\ref{s2}). Then the corresponding time
interval of the particle motion immediately follows.

The determination of the difference of
the two vectors `attached' to the nearby but distinct points is less
trivial. On one hand, this is a classical problem of the
non-euclidean geometry which is exhaustively treated in
terms of the {\it parallel transport\/} mathematically equivalent to
the notion of connection. On the other hand, such a solution would
be, in principle, a formal-mathematical one, while we need
strictly `physically meaningful' interpretations in terms of
`observable' quantities.

However one may also argue that the practice of
local measurements of spatial relationships has in fact no sufficient
experience in the dealing with the substantially curved spatial
background where the {\it spatial curvature\/} affects
the essence of the physical process realizing the measurement procedure.
Accordingly,
it cannot be \apriori\ stated quite definitely what
mathematical tools of Riemannian geometry might be
qualified as `physically realizable'. Nevertheless the parallel
vector transport along infinitesimal paths obviously {\it
may}. Indeed, the observers are certainly able to obtain the components of the
apparent metric tensor, measuring their mutual distances
by means of the light ranging. Then they may obtain the corresponding
connection (calculating the Christoffel symbols, for example) in the
point of measurement. This allows them further to determine the
corrections necessary for the re-casting the directly observable {\it
variations of components\/} of a vector under consideration into the {\it
components of\/} vector {\it variation} (represented,
mathematically, by the absolute differential contracted with the displacement
vector). Thus one may conjecture the following:
\begin{description}
 \item[(C)]
In the static apparent space, the observable variation of a vector,
dragged from a point to another nearby one, is determined by means
of the local parallel transport with respect to the
torsion-free connection compatible with the apparent metric.
 \end{description}
Said another way, living in the static curved Riemannian space, one
is to apply the corresponding standard geometric tools for the
description of the spatial relations involved in physical processes.
Anyway, such a position is the most natural and, simultaneously,
straightforward one within the framework considered.

Indeed, it would be more than strange to suppose the connection
determining `physical' parallel transport to be incompatible with
the spatial metric. Thus the only pertinent alternative could be
the introducing of a non-zero torsion. However no compelling reasons
in favor of such a modification of the statement (C) are immediately
perceptible and we limit ourselves here with these remarks.

 Nevertheless, strictly speaking, (C) is a separate additional
postulate besides the statements (A) and (B). However, one way or
another, some guiding concept has to be introduced to properly
formalize the observations over a {\it substantially curved spatial
background}. Essentially, we deal here with a sort of basic
definitions ensuring the consistent regarding of one of `primitive'
notions. Specifically, it clarifies the meaning (the method of
measurement) of a variation of a vector `dragged' in the curved space
where the existence of the preferable Cartesian coordinates yielding
trivial connection may not be stipulated. It is worth mentioning that
the conjecture (C) is the first and, in fact, the {\em only\/}
statement concerning the principles of measurements and asserting
{\em something new\/} which we use in the present work. Indeed, the
meaning of the claims~(A) and (B) (see section \ref{s2}) is, in a
sense, just opposite: essentially, they state that the space-time
curvature and acceleration of observers does {\it not\/} locally
affect the role of light signals and the basic model of time
measurements adopted in the flat space-time physics.

Ultimately,
basing on the above speculation, a straightforward calculation, whose
technical details are omitted here for brevity, yields the following
simple explicit representation of the operator of the rate of vector
variation:
$$
\bDer{}=\vc\poker\Dex,\enskip
\mbox{($\vc=\en^{-1}\mom$ is the observable velocity of a particle)}.
$$
Here $\Dex$ denotes the covariant (absolute) differential {\it in the
apparent space}, the symbol
`$\poker$' denotes the contraction of a vector (at
left) with an 1-form (at right).

It is worth emphasizing that this plausible equation is {\it not\/} a
conjectural definition of its \lhs,
\ie, a `physical derivative' of a vector
(although it would be a rather
natural one). Rather it is a straightforward consequence and follows
from the statements (A)--(C).

\section{The forces affecting the motion of test particles}

It was shown  that, utilizing the concept of the apparent space
and tracing the particle' motion, the observers reveal the
following force applied to particle:
 \begin{equation}
\fc=\en^{-1}(\mom\poker\Dex)\mom.
                                                      \label{eq8}
\end{equation}
 Using Eqs.~(\ref{eq6}), (\ref{eq5}), the \rhs\ expression can be
explicitly calculated in a general case.

Specifically, dealing with the most general case of the arbitrary
prescribed motion of a test particle (caused by the unspecified
influence on it) with arbitrary timelike worldline, the following
{\it identity\/} takes place:
 \begin{equation}
(\mom\poker\Dex)\mom\equiv\en^2\Fm+\en\crosstimes{\mom}{\Fi}+\en\fext.
                        \label{fdec}
\end{equation}
Here $\crosstimes{\cdot}{\cdot}$ denote the antisymmetric vector
cross-product in the apparent metric (\ref{eq5}). The
following series of vector fields living in the apparent space are
introduced:
 \begin{eqnarray}
\Fm&=
&\hskip 1ex\hphantom{\llap{+}}
\left[
\half{2p\P-(p^2+q^2)\P'\over(p^2+q^2)^2}
\right.
                                        \nonumber
                                                       \\
&&\left.\hphantom{\hskip 1ex\left[
\vphantom{a-\omega(a^2-p^2)\over(p^2+q^2)(\mib{P}-\mib{Q})}
                            \right.}
+
{a-\omega(a^2-p^2)\over(p^2+q^2)(\mib{P}-\mib{Q})}
\left(
\left(a-\omega(a^2-p^2)\right)
{\P'\over2\P} -2\omega p
\right)
\right]\drp
                                        \nonumber
                                                       \\
&&\hskip 1ex\llap{+}
\left[
\half{2p\Q-(p^2+q^2)\Q'\over(p^2+q^2)^2}
\right.
                                                      \label{eq10}
                                                       \\
&&\left.\hphantom{\hskip 1ex\left[
\vphantom{a-\omega(a^2-p^2)\over(p^2+q^2)(\mib{P}-\mib{Q})}
                            \right.}
-
{a-\omega(a^2+q^2)\over(p^2+q^2)(\mib{P}-\mib{Q})}
\left(
\left(a-\omega(a^2+q^2)\right)
{\Q'\over2\Q} +2\omega q
\right)
\right]\drq,
                                        \nonumber
                                                       \\
\Fi&=
&\hskip 1ex\hphantom{\llap{+}}
\left[
{2q\P\over(p^2+q^2)^2}+
{a-\omega(a^2-p^2)\over a-\omega(a^2+q^2)}\cdot
{\Q\mib{Q}'\over(p^2+q^2)(\mib{P}-\mib{Q})}
\right]\drp
                                        \nonumber
                                                       \\
&&\hskip 1ex\llap{+}
\left[
{-2p\Q\over(p^2+q^2)^2}+
{a-\omega(a^2+q^2)\over a-\omega(a^2-p^2)}\cdot
{\P\mib{P}'\over(p^2+q^2)(\mib{P}-\mib{Q})}
\right]\drq,
                                                      \label{eq10a}
\\
\en\fext&=
&\hskip 1ex\hphantom{\llap{+}}
{1\over 2(p^2+q^2)(\mom\poker d p)}
\mom\poker
\left[\P d\Cz
             \right.\nonumber\\
&&\qquad\left.
             -2\left(p^2\Ct+\Cs\right)\left(p^2d\Ct+d\Cs\right)
\right]\drp
                                                \nonumber
           \\
&&\hskip 1ex\llap{+}
{1\over 2(p^2+q^2)(\mom\poker d q)}
\mom\poker
\left[\Q d\Cz
             \right.
                                                \label{eq13}\\
&&\qquad\left.
             +2\left(q^2\Ct-\Cs\right)\left(q^2d\Ct-d\Cs\right)
\right]\drq
                                                \nonumber
           \\
&&-{1\over(p^2+q^2)}\mom\poker
\left[\hphantom{+}
{a-\omega(a^2-p^2)\over\P}\left(p^2d\Ct+d\Cs\right)\right.
                                                \nonumber
                                                          \\
&&\hphantom{-{1\over(p^2+q^2)}\mom\poker
                                       \left[
           \vphantom{a-\omega(a^2-p^2)\over\P}
                                       \right.
          }
\left.
+{a-\omega(a^2+q^2)\over\Q}\left(q^2d\Ct-d\Cs\right)
\right]\drf.
                                                \nonumber
\end{eqnarray}
It is assumed here that the worldline of the particle is timelike and,
besides, the momentum components
 $\mom\poker d p$,  $=\mom\poker d q$ are nonzero.
The latter limitation is not a substantial one and there exists
an equivalent representation of $\fext$:
 \begin{eqnarray}
  \fext=&&\mu 
 \sqrt{p^2+q^2}\sqrt{\P\Q}\sqrt{\mib{P}-\mib{Q}}
\times
       \nonumber\\
&&\hphantom{\mu}
  \left[
 \left(a-\omega\left(a^2-p^2\right)\right)\Q
   \left({d\tau\over d\nu}-p^2 {d\sigma\over d\nu}\right)
  \right.
       \nonumber\\
&&\hphantom{\mu\left[\vphantom{d\sigma\over d\nu}\right.}
\llap{$-$}
\left.
 \left(a-\omega\left(a^2+q^2\right)\right)\P
   \left({d\tau\over d\nu}+q^2 {d\sigma\over d\nu}\right)
\right]^{-1}
\times
      \nonumber\\
&&
\left\{
 {{D}\over d\nu}
 {d p\over d\nu}
                     \drp
 +{{D}\over d\nu}
 {d q\over d\nu}
                      \drq
 +\left[a(1-a\omega)
          {{D}\over d\nu}
          {d \sigma\over d\nu}
        -\omega
         {{D}\over d\nu}
         {d \tau\over d\nu}
  \right]\drf
\right\}
                                                \label{eqnn}
\end{eqnarray}
which evidently does not assert it. It is worth pointing out an additional
form of Eq.\ (\ref{eq13}) which manifests the meaning of the variation of
non-conservation of geodesic integrals (\ref{eq7}), if any:
 \begin{eqnarray}
\fext&=&
{\sqrt{\P\Q}\sqrt{\mib{P}-\mib{Q}}
        \over
(p^2+q^2)^{3/2}\left(\omega\Cs-a\left(1-\omega a\right)\Ct\right)
}\times                                         \nonumber\\
&&
\left\{
\half(\mom\poker d\Cz)\Vpq
-(\mom\poker d\Ct)\left[p^2\Vpf+q^2\Vqf\right]
\right.                                         \nonumber\\
&& \hphantom{\left\{ \half(\mom\poker d\Cz)\Vpq \right.}
\left.
+(\mom\poker d\Cs)\left[\Vqf-\Vpf\right]
\right\},
                                        \label{eqnnn}
         \\
\hbox{where}\enskip
&&\Vpq={\epsilon_{(p)}\P
        \over
        R_{(p)}
       }\drp
+
      {\epsilon_{(q)}\Q
        \over
        R_{(q)}
       }\drq,                                        \nonumber\\
&&\Vpf=
        {a-\omega(a^2-p^2)\over\P}\drf
+\epsilon_{(p)}
       {\Cs+p^2\Ct
         \over
        R_{(p)}
       }\drp,                                        \nonumber\\
&&\Vqf=
        {a-\omega(a^2+q^2)\over\Q}\drf
+\epsilon_{(q)}
       {\Cs-q^2\Ct
         \over
        R_{(q)}
       }\drq.
                                                      \nonumber
\end{eqnarray}
($R_{(p)}, R_{(q)}$ are defined by Eqs.\ (\ref{Rpq}).)
Here $\mom$ has to be interpreted as \rhs\ of Eq.~(\ref{eq6}).

The decomposition (\ref{fdec}) is derived as follows.

The contribution involving $\fext$ is defined in such a way to
accumulate all the terms
including the {\it
second order derivatives\/} of the coordinate functions specifying the
particle worldline. They are
incorporated with the corresponding expressions quadratic in the
first order derivatives in order to yield in total the (components
of) {\em covariant\/} derivative of the 4-vector
tangent to the particle worldline (\cf\ Eq.\ (\ref{eqnn})).
Said
another way, $\fext$ is chosen to represent
the constituent of the force (\ref{eq8})
causing the
{\em 4-acceleration\/} of the particle. It is \apriori\ clear that $\fext$
has to be a linear function of the components of 4-acceleration.

After subtracting $\fext$,
the residual part of \lhs\ of Eq.\ (\ref{fdec}) does not involve
second order derivatives of the coordinate functions. It depends on
their first order derivatives alone. Accordingly, the problem may be
posed to expand it with respect to the `physical basis' (\ref{eq6}).
Such a representation proves
to be {\it linear\/} (cancelling out the overall factor $\en$) with
respect to the components of momentum and the energy which are now the
only terms characterizing the details of the particle motion.
The vectors $\Fm$, $\Fi$ are just the corresponding `coefficients'
coupled with energy and momentum, respectively, independent on any
such a characteristic.

Summarizing, Eq.\ (\ref{fdec})
\begin{itemize}
 \item singles out the part of $\en^{-1}(\mom\poker\Dex)\mom$
linear with respect to the covariant derivatives of the 4-vector
tangent to particle worldline ($\fext$) and, then,
 \item
decomposes the residual part of $\en^{-1}(\mom\poker\Dex)\mom$
into the linear expansion
with the vector coefficients ($\Fm$, $\Fi$)
with respect to $\en$ and $\mom$.
\end{itemize}
The result of the above procedure is obviously unique.

The vectors $\fext$, $\en\Fm$, $\crosstimes{\mom}{\Fi}$
are the forces acting to the
test particle.  Eq.\ (\ref{fdec}) (divided to $\en$) is just
the equation of their balance. To better substantiate this statement,
let us discuss the physical aspect of the above result.

The vector $\fext$ admits a straightforward and univocal
interpretation: it represents precisely the {\it external\/} (\ie\
neither gravitational nor inertial) force applied to the massive
particle since it is immediately associated with the 4-acceleration
of particle, being a linear function of the components of the latter,
see Eq.\ (\ref{eqnn}). In particular, the 4-acceleration and $\fext$
vanish strictly simultaneously. (The direct implication trivially
follows from Eq.\ (\ref{eqnn}) while the inverse one requires
additionally the taking into account the identity
 $ g_{\alpha\beta} {d x^\alpha\over d\nu}\;
{D \over d\nu}
{d  x^\beta\over d\nu}
=0 $.)
Thus
$\fext$ vanishes if and only if the worldline is geodesic which
describes just the motion free of external non-gravitational influence.

The physical meaning of the force $\fext$ is clearly manifested in
the most simple case when the `external' influence to test particle
is caused by the electromagnetic field inherent to Kerr-Newman
space-time. To that end, let us assume that the particle carries a
small electric charge $\chrg$ whose influence to the
space-time geometry, as well as the corresponding radiative
braking, may be neglected. It is convenient to describe the motion of
such a particle in terms of `geodesic integrals' defined by Eqs.\
(\ref{eq7}). Specifically, for the free charged particle they
are not constant but
 \begin{equation}
\Cz=\cz,\enskip \Ct=\ct-(\chrg\,\Chrg){q\over p^2+q^2}, \enskip
\Cs=\cs+(\chrg\,\Chrg){q p^2\over p^2+q^2}.
                                        \label{chargedCs}
\end{equation}
Here
$\cz,\ct,\cs$ are the actual constants of motion for the
configuration in question.

Substituting expressions (\ref{chargedCs})
into Eq.\ (\ref{eqnnn}), the differentials of geodesic integrals
are calculated in explicit form. Further,
making use of Eqs.\ (\ref{endef}), (\ref{eq6}),
the geodesic
integrals are eliminated in favor of particle energy and
momentum $\en,\mom$.
Finally, the well known Lorentz' equation follows
 \begin{equation}
\fext=
\chrg(\elf+\crosstimes{\vc}{\mag}).
                                                      \label{eq11}
\end{equation}
(it is worth reminding that the speed $\vc$ equals $\en^{-1}\mom$).
Here the electric and magnetic field strengths $\elf,\mag$,
represented in the form of a complex sum $\elf+\mi\mag$, are defined
as follows:
 \begin{eqnarray}
\elf+\mi\mag&=&
-{\Chrg\over(q+\mi p)^2}
\sqrt{\P\Q\over(p^2+q^2)(\mib{P}-\mib{Q})}\cdot\Com,
                                                      \label{eq12a}\\
\mbox{where\quad}
\Com&=&
{a-\omega(a^2-p^2)\over\cal P}\drq
+\mi
{a-\omega(a^2+q^2)\over\cal Q}\drp
                                                      \label{eq12b}
\end{eqnarray}
($\Chrg$ is the charge inducing the electromagnetic field in
Kerr--Newman space-time, see \linebreak
                             Eqs.~(\ref{eq2b}), (\ref{empot})).

It is worth noting that the vectors $\elf$ and $\mag$ do {\em not\/}
depend on any characteristic of the test particle or the features of
its motion. They are determined by the properties of the space-time,
the electromagnetic field, and the rate of observers rotation
(characterized by the
formal frequency $\omega$). Thus $\elf$ and $\mag$ are just the {\it
observable electromagnetic strengths\/} detected by
uniformly rotated observers.

The above result concerning the strengths of electromagnetic field
measured by means of the analysis of motion of charged test particles
is completely concordant with what one expects basing on the
experience gained from the flat space-time electrodynamics.

Now let us consider the two further contributions to the force
balance equation (\ref{fdec}). They are associated with the
gravitational and inertial forces, the physical essence of the
configuration considered ensuring the presence of the both of them.

Indeed, let the motion of a test particle be such that the `geodesic
integrals' are constant. We have mentioned that the latter claim
corresponds to a free character of the particle motion. According to
the basic assertion of the general relativity, the free geodesic
motion of test particles is just a realization of the gravitational influence
to it which we, in turn,  associate with some gravitational force.
Additionally, one have to take into account
the effect of inertial force depending on the frame of reference
chosen, the latter force being locally inseparable from the
former one as far as one concerns their influence to particle motion. The
equation
 \begin{equation}
\fc= \en\Fm+\crosstimes{\mom}{\Fi},
                                                      \label{eq9}
\end{equation}
taking place for {\em free\/} particles describes just the influence of the
gravitational and inertial forces to them. These force a `free' particle to
deviate from a `straight line' --- a geodesic {\em in the apparent
space\/} which would be the trajectory of the particle motion,
provided the resulting force (\ref{eq9}) vanishes%
\footnote{Eq.\ (\ref{eq9}) holds true for massless test particles as
well. In particular, the photons are also subjected to the action of
the gravitational and inertial forces. This point exhibits a
substantial distinction of the present method and the approach
exploited in Ref.~\cite{ab1} (and provides us with the concise
explanation of the meaning of the well known light deflection
effect). }.
Eq.\ (\ref{eq9}) closely resembles the equation (\ref{eq11})
describing the action of electric and magnetic fields.
It is worth noting that
the `massive'
(gravitational and inertial, simultaneously) `charge' coincides with
the {\em relativistic {\rm(velocity dependent)} mass\/} of the
particle $\en$, not merely the rest mass $\mu$.

An intriguing problem is now the separating of the inertial and `true
gravitational' constituents of the combined gravitational-inertial
bundle of observable intensities (\ref{eq9}) which are, in
accordance with the principle of equivalence, linearly
superposed.

\section{Separating gravitational and inertial forces}
\label{s6}

 Dealing with forces characterizing the action of the gravitational
field, it would be important to be able to elicit their `pure
gravitational' constituents. In this case one could establish a
bridge connecting the hardly caught notion of a true gravitational
field (setting it off against inertia manifestations not associated
with `physical fields') and the quantities which can be measured in
an observation, \cf.\ \cite{MTW}. A natural methodological condition
to be \apriori\ stipulated is the operating with primordial
relativistic instruments --- watches and light rays --- in accordance
with a strictly definite relativistic measurement procedure.

There is no satisfactory universal regarding of a gravitational force
in framework of the general relativity yet. A number of different
approaches have been suggested and they lead to distinct conclusions when
treating equivalent problems (\cf\ Ref.\ \cite{grqc}). Unfortunately, no
entirely convincing criterium is now known which would enable one to
distinguish a single concept of gravitational force among suggested
ones.

Nevertheless, it should be no objections against the following
simple observation:
the gravitational field is closely connected with the space-time
curvature; in particular, it may exist (and does exist) only in the
case of a curved space-time.

A crucial role of the curvature in the qualitative recognizing of the
true gravitational field is generally accepted although the
relevant quantitative outcome obtained so far seems not so universal
and perfectly convincing. Mostly, the curvature is called to account
for the {\em tidal\/} effects caused by the inhomogeneous gravity.

We would like to suggest here a different approach which does not
pretend (in the form presented, at least) to any degree of generality
and is currently limited to the case of Kerr-Newman space-time (and,
maybe, some its nearby generalizations). Nevertheless it allows
one to describe a full scope of expectable local manifestations of
the gravitational field and reveals some its surprising properties.

The main idea realized below is straightforward: having obtained
the combined gravitational-inertial bundle of specific force intensities
$\{\Fm,\Fi\}$ determined by Eqs.\ (\ref{eq10}), (\ref{eq10a}), one may
attempt to separate in them a constituent which is intimately connected
with the curvature. One may hope that such a constituent should
describe just the a true gravitational intensities while the
residue part of $\{\Fm,\Fi\}$ is interpreted as responsible for the
effects of inertia.

At first glance such an attempt cannot allow a satisfactory
non-artificial solution. Indeed, the curvature is characterized by
mathematical expressions which necessarily involves the {\em
second\/} order derivatives of the coordinate components of the space-time
metric. On the contrary, the vector fields $\{\Fm,\Fi\}$ are
constructed from the {\em first\/} their derivatives at most.

However, the spinor components of the curvature
of the metric (\ref{eq1}) can be represented as some linear
homogeneous operators applied to the functions
 \begin{equation}
\Op=\P'(p)-{2p\over p^2+q^2}(\P-\Q),\quad
\Oq=\Q'(q)+{2q\over p^2+q^2}(\P-\Q)
                                                  \label{eq14}
\end{equation}
(\cf\ \cite{Pl}, Eqs.~(3.6) - (3.8)).
Moreover, the metric~(\ref{eq1}) is (locally) flat {\em if and only if\/}
 $$
\Op=0=\Oq.
 $$
For the sake of convenience of references, we shall call
hereinafter the expressions~(\ref{eq14}) the elements of the {\it
pre-curvature\/}%
\footnote{One may suppose that pre-curvature should be connected
with Lanczos potential (see Ref.\ \cite{lanc}) which also gives rise to
the curvature (specifically, Weyl spinor) as a result of application
of the first order differential operator. A straightforward
examination does not confirm it however.}%
 of the Kerr-Newman metric (\ref{eq1}).

Thus although the curvature involves the
second order derivatives of the metric components (in our case the
second order derivatives of the functions $\P$ and $\Q$), it also may
be interpreted as a subsidiary object derived from pre-curvatures
(\ref{eq14}). In a sense, $\Op,\Oq$ may be used as the equivalent
representation of the curvature.

The above property of the curvature of the Kerr-Newman space-time
and the explicit mathematical structure of the
expressions of $\Fm, \Fi$, see Eqs.~(\ref{eq10}), suggest the
following natural statement realizing the `minimal coupling' of the
`true gravity' and the curvature:
 \begin{description}
\item[Conjecture:]
{\sl In the case of metric~\hbox{\rm(\ref{eq1})} the true
intensities of the gravitational field are linear homogeneous functions
of the pre-curvature elements\/}~(\ref{eq14}).
 \end{description}
Indeed, it is impossible to separate in \rhs's of Eqs.~(\ref{eq10})
the other additive sub-expressions which would vanish {\em strictly
simultaneously\/} with the pre-curvatures (and the curvature itself)
{\it for arbitrary\/}
functions $\P$ and $\Q$.

Then, adopting the validity of the above conjecture, a
straightforward inspection of Eqs.~(\ref{eq10}), (\ref{eq10a})
immediately
enables one to find the desirable {\it gravitational attractive\/}
(gravitoelectric) and {\it gyroscopic\/} (gravitomagnetic)
intensities $\Fgr$ and $\Fgyr$. The most advantageous way of their
representation turns out to be
the introducing of the complexified vector sum
$\Fgr-\Ratio{i}{2}\Fgyr$ (similar to the complex vector $\elf+i\mag$
utilized in electrodynamics) which proves to be factorizable as follows:
 \begin{equation}
\Fgr-\Ratio{\mi}{2}\Fgyr=
-\half{
\left(a-\omega\left(a^2+q^2\right)\right)\Oq
-\mi
  \left(a-\omega\left(a^2-p^2\right)\right)\Op
\over
(p^2+q^2)(\mib{P}-\mib{Q})}\cdot\Com.
                                                     \label{eq15}
 \end{equation}
The complex-valued vector field $\Com$ has been defined in
Eq.~(\ref{eq12b}).

The residue parts of the specific forces $\Fm$ and $\Fi$ which leave
after the subtracting of $\Fgr$ and $\Fgyr$, respectively, provide
obviously the {\it inertial centrifugal\/} $\Fcf$ and {\it
Coriolis\/} $\Fcor$ intensities. They are equal to
 \begin{eqnarray}
\Fcf&=&
{\omega^2\over\mib{P}-\mib{Q}}
\left[q\drq-p\drp\right],
                                                \label{eqcf}
                                     \\
\Fcor&=&
{2\omega\over(p^2+q^2)(\mib{P}-\mib{Q})}
\left(
{a-\omega(a^2-p^2)\over\cal P}
-
{a-\omega(a^2+q^2)\over\cal Q}
\right)\times
                                                \label{eqcor}
                                        \nonumber\\
&&%
\hphantom{\omega^2\over\mib{P}-\mib{Q}}
\left[q\P\drp+p\Q\drq\right]
                         \label{eq16}
\end{eqnarray}
and determine jointly the {\it inertial force\/}
$$
\fir=\en\Fcf+\crosstimes{\mom}{\Fcor}.
$$

Examining the results obtained, it is worth stressing that the
remarkable proportionality ({\em generalized alignment}) of the
`complexified gravitational intensity'~(\ref{eq15}) and the `complex
electromagnetic strength'~(\ref{eq12a}) was in no way assumed initially.
It is therefore a surprising and highly promising fact. Although
the implications of the coincidence of `complexified directions'
of the gravitational and electromagnetic field strengths in the
Kerr-Newman space-time are not manifest yet, this relationship cannot be
a mere occasion. In any case, it should be regarded as a strong
indirect argument in favor of the plausibility of the expressions for
the true gravitational intensities suggested, as well as gives evidence
for the validity of the approach applied at whole.

Another curious fact worth mentioning here is the orthogonality of
the centrifugal and Coriolis intencities: $\Fcf\cdot\Fcor=0$. It also was
not assumed beforehand, gaining therefore some new insight into the
properties of the inertia manifestations.

\section{Boyer-Lindquist picture}
\label{s8}

It is worth transforming the main relationships derived above to the
Boyer-Lindquist coordinates \cite{BL} 
which are usually
utilized when considering the physics-related properties of Kerr and
Kerr-Newman metrics, \cf\ Ref.\ \cite{ON}.

Reminding the standard abbreviating notations
 $$
\Delta=\Delta(r)=r^2-2mr+a^2+\Chrg^2,\quad
\rho^2=\rho^2(r,\theta)=r^2+a^2\cos^2\theta
$$
and introducing the following additional ones
 \begin{eqnarray}
\Up&=&\Up(\theta)=1-\omega a\sin^2\theta,
                                                      \label{zom}
                                                        \\
\Uq&=&\Uq(r)=a-\omega(a^2+r^2),
                                                        \\
\Gamma&=&\Gamma(r,\theta)=
\Up^2\Delta-\Uq^2\sin^2\theta>0,
                                                        \\
\Comm&=&
\Up\Delta\drr-\mi\Uq\sin^2\theta\drt,
                                                \label{comm}
\end{eqnarray}
it can be shown that the complexified strength of the electromagnetic
field $\elf+i\mag$, the complexified intensity of gravitational
field $\Fgr-\Ratio{i}{2}\Fgyr$, and and intensities of
inertia manifestations $\Fcf, \Fcor$ equal
 \begin{eqnarray}
\elf+\mi\mag&=&
-{\Chrg\over(r+\mi a\cos\theta)^2}
{\Comm\over \sqrt{\rho^2\Gamma}},
                                            \label{ems}
                                                        \\
\Fgr-\Ratio{\mi}{2}\Fgyr&=&
-
\left\{
\left(m\left(r^2-a^2\cos^2\theta\right)-\Chrg^2 r\right)\Up
                                                \right.\nonumber\\
&&\hphantom{-\left\{%
\vphantom{\left(m\left(r^2-a^2\cos^2\theta\right)\right.}%
\right.}\left.
+\mi\left(2mr-\Chrg^2\right)\Uq\cos\theta
    \right\}
{\Comm\over\rho^4\Gamma},
                                        \label{fgrgyr}
                           \\
\Fcf&=&
\omega^2
{\Delta\sin\theta
        \over
\Gamma}
\left[
r\sin\theta\drr+\cos\theta\drt
     \right],
                                        \label{fcf}
                            \\
\Fcor&=&
2\omega
{\Up\Delta-a\Uq\sin^2\theta
        \over
\rho^2\Gamma}
\left[
\Delta\cos\theta\drr-r\sin\theta\drt
     \right].
                                        \label{fcor}
\end{eqnarray}
It is worth noting that except of the $\Gamma$-dependent factors
involved in denominators (which play the role of the special relativistic
$\gamma$-factor) all the dependencies on the formal frequency of
observers rotation $\omega$ are either {\em linear\/} (for
electromagnetic field) or {\em quadratic\/} (for gravitational field
and inertia).

The representation (\ref{zom})-(\ref{fcor}) is convenient in
particular for the analysis of the physically important limiting
cases. Choosing $a=0$, the field strengths in the
Reissner-Nordstr\"om space-time follow. If $\Chrg=0$ the
characteristics of the Kerr field and, with $a=0$, Schwarzschild field
immediately result.

A weak field limit can be easily examined as well. It can be shown
that it completely conforms to the corresponding non-relativistic
(Newtonian) values. Additionally, the following expression for the
gyroscopic (gravitomagnetic) gravitational intensity detected by
asymptotically {\it non-rotated\/} observers ($\omega=0$) in the case
of Kerr space-time is of interest:
 \begin{equation}
\Fgyr\simeq{2ma\over r^3}
\left(2\cos\theta\drr-{\sin^2\theta\over r}\drt\right).
                                          \label{kerrgyr}
\end{equation}
It possesses no immediate Newtonian analogue, of course.

\section{Summary and discussion}\label{sum}

Summarizing,
we started with a simple observation that in the case of the
stationary Kerr-Newman space-time the natural assumption on the local
constancy of the light speed (statement (A) of section \ref{s2}) and
the well known description of ideal clocks (statement (B) therein)
imply the unique manifest representation of the spatial relations
referring to the notion of a distance.
Specifically, the uniformly rotated observers are in fact
immersed in the `constant' (static) curved Riemannian space endowed
with the metric (\ref{eq5}) --- the `apparent space'.

Accordingly, it immediately follows from the above statement that,
tracing the motion of a test particle, the observers find that its
vector velocity, vector momentum and energy are described by the
equations (\ref{eq6}) which are deduced by means of the standard
relativistic mechanics applied in vicinity of the observation
point.

Next, given the energy and momentum of a particle, their
rate of variation can be determined, bearing in mind the subsequent
introducing
the notion of observable force acting to a particle.
At this stage
an additional
guiding assumption is to be postulated: it asserts that the
infinitesimal variation of a vector `attached' to the moving particle
is determined with the help of the torsion-free connection compatible
with the apparent metric which yields the `physical measure'
of a distance (see the claim (C) in section \ref{s4}).

After some technical work we obtain the explicit expression
$\en\Fm+\crosstimes{\mom}{\Fi}+\fext$ for the observable force $\fc$
applied to the particle (for the meaning of the notations see
Eqs.~(\ref{eq10}), (\ref{eq10a}), (\ref{eq13})--(\ref{eqnnn})).

The principle of the separating the above constituents of the total
force $\fc$ determines their meaning. Specifically, $\fext$ is the
external (non-gravitational and non-inertial) force closely connected
with the standard 4-force (its components are linear homogeneous
functions of the components of the latter). In the case of a free
charged particle it coincides with the Lorentz force
$\chrg(\elf+\crosstimes{\vc}{\mag})$, where $\elf$ and $\mag$ are the
observable electric and magnetic strengths, respectively, determined
by Eq.~(\ref{eq12a}). Further, $\en\Fm$ is the attractive `massive'
mixed gravitational-inertial force, $\Fm$ being characteristic of its
intensity. Similarly, $\crosstimes{\mom}{\Fi}$ is the mixed
gravitomagnetic-Coriolis force possessing the own intensity $\Fi$, It
is important to note that both $\Fm$ and $\Fi$ do {\em not\/} involve
any specific characteristic of a test particle or its motion and
are determined by the space-time geometry (and depend on the
observers).

The final crucial step enabling one to describe in physical terms the
gravitational field associated with the model considered is a
plausible separation of the combined gravitational-inertial forces
(and the corresponding intensities) into their purely gravitational
and purely inertial constituents. The trick utilized for this purpose
is based on the inspection of the variation of the combined
gravitational-inertial force `in response' to the variation of the
space-time geometry running over the class of metrics described by
Eq.~(\ref{eq1}) which involves two arbitrary functions ${\cal P}(p)$
and ${\cal Q}(q)$. (Strictly speaking, this procedure requires the
leaving the class of Kerr-Newman metrics. Said another way, it would
be insufficient to take the metric (\ref{eq1}) with the functions
${\cal P}(p)$, ${\cal Q}(q)$ fixed to the form (\ref{eq2b}) or
(\ref{eq2a}) alone.)

More definitely, the common terms in the gravitational-inertial
intensities and the space-time curvature components which vanish {\em
strictly simultaneously for arbitrary functions\/} $\cal P$, $\cal Q$ are
singled out. (In accordance with their role these terms
were named pre-curvatures, see section~\ref{s6}). There exist all the
reasons to identify the corresponding constituents of the intensities
with the {\it pure gravitational contribution}. Taking them off from
the combined gravitational-inertial forces, it is natural to
interpret the residual forces as {\it pure inertial\/} in a nature.

It is worth noting that the method of the separating the inertial and
gravitational intensities applied --- a `variation' of the space-time
geometry --- cannot be associated with any conceivable `local
experiment' and thus our result does {\it not\/} conflict with the
equivalence principle (\cf\ Ref.\ \cite{FF2}).

Ultimately, we obtain the attractive gravitational (gravitoelectric)
intensity $\Fgr$ and the gyroscopic (gravitomagnetic) intensity
$\Fgyr$ determined by Eq.~(\ref{eq15}) and Eq.~(\ref{fgrgyr}), the
centrifugal inertial intensity (Eqs.~(\ref{eqcf}), (\ref{fcf})), and
the Coriolis intensity (Eqs.~(\ref{eqcor}), (\ref{fcor})),
respectively.

These equations reveal a proper behavior in the weak field limit.
Moreover, the tools which we used for their derivations are based
on a remnant of the model which is directly inherited from the
physics of Minkowski space-time. Accordingly, the method applied
proves to be strictly unique and gains advantage of an unambiguous
physical interpretation.

 It is instructive to mention that the specific role of light signals
(expressed by the statement (A) in section (\ref{s2})) does not
attribute to the trajectories of light rays the role of a `standard'
of a constant direction. Specifically, we used the statement (A) for
the deriving the formula (\ref{eq5}) determining local distance. Here
the consideration is local and the geodesic property of worldlines of
photons (second order effect) is {\em not\/} actually used. All what
we need here is the vanishing on them of the interval (\ref{eq1}),
\ie\ the fact that worldlines of light signals are null. Accordingly,
there is no reason for finite segments of light rays to be straight
(geodetic) with respect to the observable geometry of the apparent
space. It is indeed the case and, generically, the {\em observable
acceleration\/} of photons does not vanish. In physical terms this
means just the varying of the ray direction, providing the
theoretical base for invariant interpretation of the light deflection
effect.

Photons feel the same gravitational and inertial forces as the
particles possessing a rest mass do. However the extrinsic force
$\fext$ is here identically zero (precisely because photon worldlines
are geodesic). In this respect our approach differs from one by
Abramowicz {\it et al\/} where the rays are regarded as `dynamically
straight' \cite{ab1}.

 Another point worth mentioning is the problem of a kind of mass
involved in gravitational interaction. Here we shall speak about the
mass of test particles. In according with the above derivation their
`gravitational charge' is $\en=\mu/\sqrt{1-\vc^2}$ (let us remind
that the scalar product and, in particular, $\vc^2$ is to be
determined with respect to the metric of the apparent space). Thus we
have an instance of manifestation of the equivalence principle
claiming in our case that the inertial mass $\mu/\sqrt{1-\vc^2}$
characterizes also the gravitational properties of a particle.

Now we would like to mention some simple consequences
which are more or less automatically inferred from the results
presented.

 First, it is worth noting that the formulae (\ref{ems}),
(\ref{fgrgyr}) yield certain characteristic of the `classical spin'
of the electromagnetic and gravitational fields. Obviously, it can be
associated with the power of the factor $\Gamma^{-1/2}$, \cf\
\cite[Eq.\ (12a)]{FF2}. Hence, measuring strengths of the
electromagnetic field for the various angular velocities of
observers, one finds the spin equal to 1 while for the gravitational
field it amounts to 2. One thus has to be careful when resorting to
too direct analogues between the gravitation and the
electromagnetism.

 Further, although it could seem surprising, one has to doubt of the
widespread regarding of the gravitational field as a source of tidal
effects, assuming it to be proportional to (the proper projections
of) the curvature tensor. Indeed, we see that in the case of
Kerr-Newman field at least {\em both\/} the curvature and field
intensities are subsidiary objects derived from another one named
pre-curvature which depends on the {\em first\/} order derivatives of
the metric. (Unfortunately, the covariant meaning of pre-curvature is
not known yet.) It is essential that the curvature involves the
derivatives of the pre-curvature elements while the gravitational
field intensities are simply their linear functions. Thus {\em the
intensity of gravitational field is not a derivative of the
space-time curvature.} They are not algebraically connected (but
nevertheless vanish strictly simultaneously).

Finally, one may express a hope that the surprising remarkable
`complex proportionality' of `the complexified strengths'
$\elf+\mi\mag$ and $\Fgr-\Ratio{\mi}{2}\Fgyr$ for the electromagnetic
and gravitational fields, respectively (see Eqs.~(\ref{eq12b}),
(\ref{comm})), as well as the very singling out the purely
gravitational field intensities, could yield a new insight in our
concepts concerning the physical manifestations of the gravitational
field.


\end{document}